# IEEE 1149.4 compatible ABMs for basic RF measurements

Pekka Syri[1], Juha Hakkinen[1] and Markku Moilanen[2]

Department of Electrical and Information Engineering, Electronics Laboratory[1], Optoelectronics and Measurement Techniques Laboratory[2], University of Oulu, Finland

**Abstract**

*An analogue testing standard IEEE 1149.4 is mainly targeted for low-frequency testing. The problem studied in this paper is extending the standard also for radio frequency testing. IEEE 1149.4 compatible measurement structures (ABMs) developed in this study extract the information one is measuring from the radio frequency signal and represent the result as a DC voltage level. The ABMs presented in this paper are targeted for power and frequency measurements operating in frequencies from 1 GHz to 2 GHz. The power measurement error caused by temperature, supply voltage and process variations is roughly 2 dB and the frequency measurement error is 0.1 GHz, respectively.*

## 1 Introduction

An emerging testing standard IEEE 1149.4 allows the access of selected locations in the ASIC for testing purposes [1]. The standard uses so-called analogue boundary modules (ABM) to access the internal and external signals of the ASIC. However, the standard is mainly targeted for low-frequency testing, wherefore it is not directly applicable to high-frequency use.

In order to allow the measurement of RF signals, some kind of signal processing must be performed inside the ABM. This processing extracts the information one is measuring from the RF signal and represents the result as a low-frequency signal, which can be further processed by structures specified in the standard (1149.4). Specifically, the aim of this study is to design the required RF analogue circuit structures (i.e. RF-to-LF signal processing) needed to produce IEEE 1149.4 compatible ABMs for RF frequency and power measurements.

## 2 Test circuit

The circuit is composed of two individual ABM structures, the first being a basic RF-ABM shown in Fig. 1 and the other containing preamplifiers, which allows the measurement of weaker signals. The main parts of the basic RF-ABM shown in figure 1 are frequency (Fdet) and power detectors (Pdet) which output DC quantities relative to the input signal power and frequency, respectively.

Measurement results (Vout, out- and out+) from and tuning inputs (tunef and tuneP) to the detectors themselves can be connected to the IEEE 1149.4 type analogue test port (ATP) using a programmable switch matrix (.4 MUX). Both ABM structures are controlled with an external control unit (PC, for example) using a serial data bus (signals labelled *select* in the figure originate from this serial data).

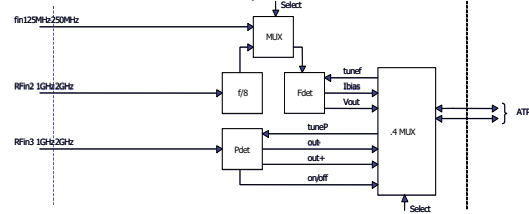

**Fig. 1. The basic ABM.**

### 2.1 Power detector

Literature reports many possible circuit topologies, which may be used for RF power measurement [2][3][4]. Unfortunately the circuit process, which was used to realize the ABM test circuit, did not offer support for diode- or BJT-based power detectors. Therefore, a circuit structure comprising only a MOS transistor shown in Fig. 2 was developed.

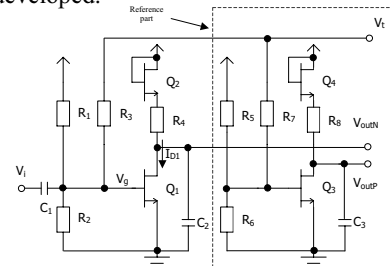

**Fig. 2. MOS transistor based RF power detector.**

In this circuit, the gate of transistor $Q_1$ is biased exactly to the threshold voltage, external tuning of this voltage is possible through the IEEE 1149.4 analogue bus via pin $V_t$. Now, $Q_1$ will conduct current $I_D$ only for the positive half cycles of the sinusoidal input voltage $V_i$, thus creating an output voltage across the load composed of $R_4$ and $Q_2$ similar to that of the half wave rectifier. The DC value, which is relative to the input power of this output signal, is extracted by the low pass filter formed by the load and $C_2$. Output voltage of the power detector can be calculated by equation [5]

$$V_{out} = V_{outN} - V_{outP} = -I_{DC}R_4 - \sqrt{\frac{2I_{DC}}{K\frac{W_2}{L_2}}}, \quad (1)$$

where $I_{DC}$ = DC value of the half wave rectified drain current of Q1.

### 2.2 Frequency detector

The frequency detector shown in Fig. 3 was used in the test circuit. The operation of the circuit is depicted in reference [6].

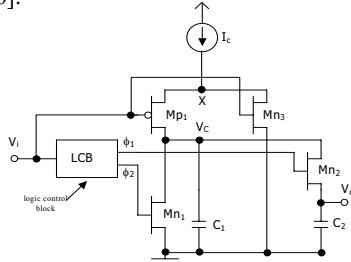

**Fig. 3. The schematic of the frequency detector.**

Main operating principle of the frequency detector is that after many periods of the input signal, the voltage across $C_2$ will reach a stable (DC) value, which is related to the input frequency $f$ (=1/T) by equation [6]

$$V_o = \frac{I_c}{C_1}\left(\frac{T}{2}\right) = \frac{I_c}{C_1}\left(\frac{1}{2f}\right). \qquad (2)$$

## 3 Measurement results

Both ABM structures were tested to ensure they are operating according to the simulations. More accurate characterization measurements were performed for the basic ABM only. ABM structures were DC-calibrated before measurements using tuning connections (tuneP and tunef) [see fig 1.].

According to measurements made on the silicon, the accurate measurement range of the power detector is from 1.2 GHz to 1.8 GHz.

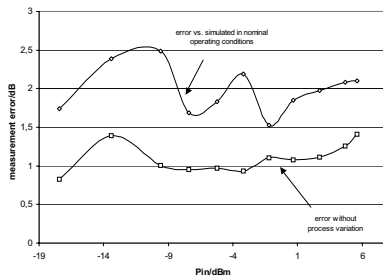

**Fig. 4. Power measurement error vs. simulated response and error without process variation.**

Figure 4 shows the calculated power measurement error vs. the simulated response and the error without process variation. In these measurements the frequency of the measured signal was 1.5 GHz (the middle frequency of the accurate measurement range), the supply voltage was 2.5 V±0.25 V and the temperature range was –10 °C…70 °C. According to the measurement results the power measurement range of the basic ABM is –18 dBm…6 dBm. The preamplifiers of the other ABM increase the measurement range to the lower power levels and the final range is from –25 dBm to –3 dBm.

Figure 5 shows the calculated frequency measurement error vs. the simulated response and the error without process variation (the supply voltage was 3.3 V±0.3 V and the temperature range was –10 °C…70 °C). According to the measurements the power level of the measured signal for the frequency detector of the basic ABM has to be at least +5 dBm and the representative value for the structure including preamplifiers is –5 dBm.

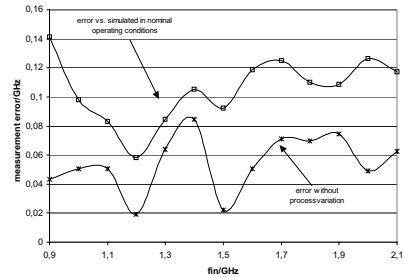

**Fig. 5. Frequency measurement error vs. simulated response and error without process variation.**

## 4 Conclusions

It seems according to the simulations and measurements that it is possible to use the IEEE 1149.4 testing standard for radio frequency testing. The ABMs developed in this study can be used in basic frequency and power measurements. DC-calibration developed in this study decreases measurement errors considerably. The power measurement error caused by the temperature, supply voltage and process variations is roughly 2 dB and the frequency measurement error is 0.1 GHz, respectively. If the error caused by the process variation is calibrated out, the power measurement error decreases to 1 dB and the frequency measurement error to 0.05 GHz, respectively.